\documentclass[10pt]{article}
\usepackage{amsmath}
\usepackage{amssymb}
\usepackage{graphics}
\usepackage{rotating}
\usepackage{cite}
\usepackage{color}
\usepackage{colortbl}
\usepackage{fancybox}
\usepackage{pstricks}
\usepackage{color, colortbl}

\definecolor{bgA}{rgb}{0.98, 0.93, 0.36}
\definecolor{ay}{rgb}{0.91, 0.84, 0.42}
\definecolor{arsenic}{rgb}{0.23, 0.27, 0.29}
\definecolor{darkseagreen}{rgb}{0.21, 0.37, 0.23}
\definecolor{darkred}{rgb}{0.55, 0.0, 0.0}
\definecolor{asparagus}{rgb}{0.53, 0.66, 0.42}
\definecolor{med}{rgb}{0.79, 0.86, 0.54}
\definecolor{arylideyellow}{rgb}{0.91, 0.84, 0.42}
\def\0{\mbox{\tiny $0$}}
\def\1{\mbox{\tiny $1$}}
\def\2{\mbox{\tiny $2$}}
\def\3{\mbox{\tiny $3$}}
\def\4{\mbox{\tiny $4$}}
\def\5{\mbox{\tiny $5$}}
\def\6{\mbox{\tiny $6$}}
\def\7{\mbox{\tiny $7$}}
\def\8{\mbox{\tiny $8$}}
\def\9{\mbox{\tiny $9$}}
\title{\hspace*{-1.2cm}\shadowbox{\fcolorbox{black}{ay} { {\color{arsenic}{ \large \bf \begin{tabular}{c}
WEAK MEASUREMENT OF THE COMPOSITE\\ GOOS-H\"ANCHEN SHIFT  IN THE CRITICAL REGION.
\end{tabular}}}}}}

\author{
\small  Oct\'avio J. S. Santana$^{\1}$,\,\, Silv\^ania A. Carvalho$^{\2}$,\,\, Stefano De Leo$^{\3}$\thanks{deleo@ime.unicamp.br}\,\,\,\,and\,\,Lu\'is E. E. de Araujo$^{\1}$ \\
\small $^{\1}$ Institute of Physics ``Gleb Wataghin'', University of Campinas (UNICAMP), Campinas, S\~{a}o Paulo, Brazil \\
\hspace{-1.2cm}\small $^{\2}$ Department of Mathematics, Physics and Computation, Universidade do Estado do Rio de Janeiro (UERJ), Resende, Brazil \\
\small $^{\3}$ Institute of Mathematics, Statistics and Computing Science, University of Campinas (UNICAMP), Campinas, Brazil\\}

\date{\small
\fcolorbox{black}{med} {\color{darkred} $\bullet$ {\color{arsenic}{
{\small \bf  Optics Letters {\color{darkred} 41}, 3884-3887 (2016)
}}} {\color{darkred}{ $\bullet$}} } }

\begin{document}


\maketitle

\vspace*{-.7cm}

\begin{abstract}
\noindent
By using a weak measurement technique, we investigated the interplay between the angular and lateral Goos--H\"anchen shift of a focused He-Ne laser beam for incidence near the critical angle.  We verified that this interplay dramatically affect the composite Goos-H\"anchen shift of the propagated beam. The experimental results confirm theoretical predictions that recently appeared in the literature.
\end{abstract}







\section*{\color{darkred} \normalsize I. INTRODUCTION}
Finite optical beams do not always follow the laws of geometrical optics. A beam of light, after reflecting from a planar interface, will shift from the path predicted by ray optics. The shift may occur in a direction that is either parallel or perpendicular to the plane of incidence, corresponding to the Goos-H\"anchen (GH) and Imbert-Fedorov (IF) shifts, respectively~\cite{Rev2}. For incidence angles below the critical angle $\theta_c$ for total internal reflection (TIR), under partial reflection, the shift corresponds to an angular deflection of the beam, and it exhibits an axial dependance consisting of a linear function of beam propagation distance. Above $\theta_c$, under TIR, the beam shift is lateral in nature, and it shows no axial dependence for a well collimated beam. But for a focused incident beam, the GH and IF shifts may become linearly dependent on the beam propagation distance, leading to a propagation enhancement of both shifts~\cite{Role}. This linear enhancement due to propagation was experimentally observed in~\cite{2008Sci}.

The lateral GH shift has its physical origin in the angular dispersion of the complex Fresnel reflection coefficients.  Its angular counterpart results from Fresnel filtering~\cite{Rev2}, which leads to a spatial asymmetry, or deformation, of the optical beam. From a theoretical point of view, the study of laser propagation in dielectric blocks plays an important role in understanding under which conditions it is possible to observe the development of an asymmetry in optical beams~\cite{2013JMO}. Optical beam shifts are interesting not only because of their fundamental nature, but also because of their applications in areas such as optical microscopy~\cite{MicroOptica}, high-sensitivity temperature sensors~\cite{TempSens1} and high-sensitivity chemical vapor detection~\cite{VapDet}.

For a finite beam with waist radius $w_0$, the transition from an angular to a lateral GH shift occurs in the critical region~\cite{ana2}
\begin{equation}
\theta_c -\frac{\lambda}{w_{0}} < \theta <  \theta_c +\frac{\lambda}{w_{0}},
\end{equation}
where $\theta$ is the angle of incidence, and $\lambda$ is the wavelength. For a very large beam, there is no critical region, and the transition from angular to lateral shift occurs at $\theta_c$. However, for a small $w_0$, there exists a range of incidence angles for which the beam experiences a composite GH (CGH) shift, consisting of both an angular and a lateral shift~\cite{1987JOSAA,2006EURO}. The GH shift has been studied theoretically near $\theta_c$~\cite{ana2,1987JOSAA,1986JOSAA,1988JOSAA,ana1,2015JO}. It has been shown that the shift depends on $w_0$ and $\theta$, and it peaks at an angle $\theta$ slightly larger than $\theta_c$. But to this date, there is limited experimental data available in the critical region, taken in the microwave~\cite{1977JOSA, 2006EURO, 2011NEWJNL} and mid-infrared~\cite{1992PRL} regimes. Furthermore, theoretical and experimental results for near-critical incidence are not in good agreement~\cite{Rev2}. Therefore, new experimental and theoretical work in the critical region are necessary to better elucidate the physics of the GH effect near critical incidence.

Theoretical studies~\cite{1987JOSAA,2015JO} have shown that the interplay between the angular and lateral GH shifts in the critical region will greatly affect the CGH shift of a propagated beam. In Ref.~\cite{1987JOSAA}, an approximate analytical expression for the CGH is derived, showing that this interplay results in a {\em nonlinear} dependence of the CGH shift on the beam propagation distance $z$. A recent work by one of the authors~\cite{2015JO} showed that the nonlinear axial $z$ dependance leads to an enhancement of the CGH shift for a propagated beam with respect to the shift evaluated at the near field ($z \ll k w_0^2$, where $k =  2\pi / \lambda$).

Lateral beam shifts are very small (typically on the order of $\lambda$) compared to the physical size of the beam. And angular shifts are usually smaller than the beam divergence. Therefore, beam shifts are difficult to measure for optical wavelengths. To overcome this difficulty, an optical analog of the weak measurement technique \cite{2012NJP14} has been successfully used to investigate the lateral~\cite{2013OL} and angular~\cite{2014OLa} GH shifts, as well as the IF shift~\cite{2014OL}. In a weak measurement, the measured system is projected onto a certain post-selected final state, nearly orthogonal to the initial state, giving rise to a measured weak value that may take on very large (amplified) values. The measurement is weak in the sense that the magnitude of the beam shift is much smaller than the beam diameter. In \cite{2013OL}, using weak measurements, Merano and colleagues experimentally verified, in the near infrared, the classical Artmann result \cite{Rev2} for incidence angles far above $\theta_c$. An interesting feature of their result is that the amplification of the GH shift was a constant factor of 100 for any angle of incidence.

\section*{\color{darkred} \normalsize II. EXPERIMENTAL SETUP}

In this Letter, we use a weak measurement scheme to experimentally observe the CGH shift, and we test the theoretical predictions given in~\cite{2015JO} regarding the effect of beam propagation on the CGH shift. Ours is also the first experimental investigation of the CGH shift with visible light.

\begin{figure}[h]
\hspace*{0.2cm}
\includegraphics[scale=2.0]{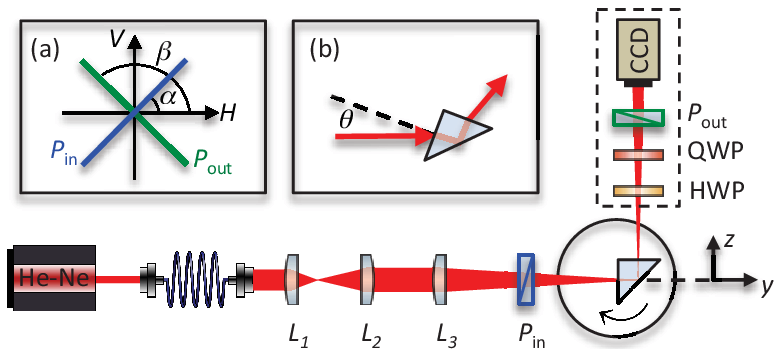}
 \caption{Experimental setup of the optical weak measurement experiment for observation of the Goos H\"anchen shift in the critical region. $L_{1,2,3}$ are lenses; HWP (QWP) is a half (quarter) wave plate; and $P_{\rm{in,out}}$ are linear polarizers. The QWP, HWP, $P_{\rm {out}}$ and CCD camera are mounted on a common platform (dashed rectangle). Inset (a) shows the orientation of the transmission axis of $P_{in}$ and $P_{out}$ relative to the horizontal ($H$) and vertical (V) axes. The $H$ ($V$) axis is parallel (perpendicular) to the optical beam's plane of incidence on the prism. As shown in inset (b), the angle of incidence $\theta$ is measured relative to normal incidence on the prism's front leg; a negative $\theta$ corresponds to a clockwise rotation of the prism. The origin of the $yz$ coordinate system is located at the point where the focused beam hits the prism's hypotenuse.}
\label{fig:Fig1}
\vspace*{1cm}
\includegraphics[width=\linewidth,height=6cm]{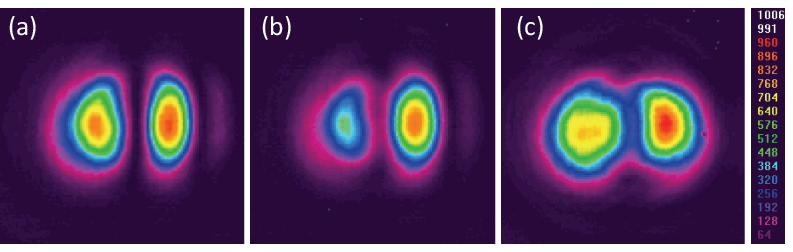}
 \caption{Measured beam profiles for (a) $\theta = -5.55^{\circ}$ ($\Delta\epsilon = 0$), (b) $\theta = -5.55^{\circ}$ ($\Delta\epsilon = 0.01$), and (c) $\theta = -5.60^{\circ}$ ($\Delta\epsilon = 0.01$).}
\label{fig:Fig2}
\end{figure}

Figure~\ref{fig:Fig1} illustrates our experimental setup. The output of a He-Ne laser (at $\lambda = 632.8$~nm) is coupled into a single mode fiber for spatial filtering purposes as well as for improving the beam-pointing stability. The output beam of the fiber is expanded and collimated with lenses $L_{1}$ and $L_{2}$. The collimated  beam passes  through lens $L_{3}$ (with an effective focal length of 1~m) producing a focused  beam with beam waist radius $w_0 = (169.5 \pm 0.5) \, \mu$m. The focused beam can be considered to be Gaussian to a very good approximation given its measured $M^2 = 1.075 \pm 0.005$. Polarizer $P_{_{\rm in}}$ and analyzer $P_{_{\rm out}}$ are nanoparticle linear film polarizers (extinction ratio of 100,000:1) used to pre- and post-select the linear polarization state of the light at angles $\alpha$ and $\beta$, respectively, defined in the inset (a) of Fig.~\ref{fig:Fig1}. We orient $P_{_{\rm in}}$ such that $\alpha = \pi/4$. For this polarizer angle, the beam incident on the prism has both $s$ and $p$ linear polarization components. The  $45^{\circ}-90^{\circ}-45^{\circ}$ prism has a 12.5~mm leg size. It is made of BK7 glass with  a refractive index of $n = 1.515$ at 632.8~nm, so the corresponding critical angle for TIR is $41.30^\circ$. The prism legs are anti-reflection coated to reduce unwanted internal reflections, but its hypotenuse is uncoated. The prism is mounted on a  high-precision rotation mount equipped with a micrometer and vernier scale that together provide a 5 arcmin resolution when rotating the prism to adjust the beam's angle of incidence. The high resolution of the rotation mount is an important aspect of the experimental setup since
the critical region is very narrow: $\lambda/w_0 \approx 0.2^{\circ}$. The prism's front leg is centered on the rotation axis of the rotation mount such that the incidence angle $\theta$ is measured relative to normal incidence on the front leg, as illustrated in inset (b). With this definition, the critical angle occurs at $\theta_{c} = -5.603^{\circ}$. The laser beam is incident on the front leg of the prism, reflects at its hypotenuse (which is positioned at the beam waist of the focused laser beam), and exits via the prism's back leg. After reflection, the beam propagates along the $z$ axis. Rotatable zero-order half (HWP) and quarter (QWP) wave plates are used to cancel the relative phase difference between the orthogonal $s$ and $p$ polarization components acquired by the beam upon reflection~\cite{2013OL}. Analyzer $P_{_{\rm out}}$ is mounted on a high-precision rotational mount that allows us to set its orientation with a resolution of 5 arcmin. After emerging from $P_{_{\rm out}}$, the diverging laser beam is detected by a CCD camera. The camera monitors the beam spatial profile, which is analyzed by a beam-pointing stability software to measure the beam centroid position with a typical $5\,\mu$m precision. The wave plates, analyzer and CCD camera are mounted on a platform that is repositioned whenever the prism is rotated so that the laser beam always impinges perpendicularly onto the camera.

To measure the CGH shift, associated with the $y$ coordinate, we follow a similar procedure to that prescribed in~\cite{2013OL} for a weak measurement: For any particular angle of incidence, with polarizer $P_{_{\rm out}}$ set to orthogonal position relative to $P_{_{\rm in}}$ ($\beta = \alpha + \pi/2$), we  adjust the QWP and HWP in order to minimize the transmitted power. Figure~2a shows a typical resulting transmitted beam profile. It consists of two peaks separated by $2^{1/2} w(z)$, where $w(z)$ is the beam $1/e^2$ radius expected for the input Gaussian beam at the CCD plane. Next, we turn $P_{_{\rm out}}$ to  $\beta = \alpha + (\pi/2) + |\Delta\epsilon|$, where
$\Delta\epsilon \ll 1$. Figure~2b shows the observed beam profile. Far above $\theta_c$, one would observe a single Gaussian peak, as reported in~\cite{2013OL}. But near $\theta_c$ we see that the beam profile consists of two peaks, one being much larger in intensity than the other. In our beam pointing stability software we select a region of interest containing the larger peak and measure its centroid position. The software acquires data at a fast rate, quickly gathering 1000 data values for the centroid position. We then set $P_{_{\rm out}}$ to  $\beta = \alpha + (\pi/2) - |\Delta\epsilon|$ and again measure the centroid position of only the larger peak. Having measured the two centroids positions, we determine their separation $\Delta y_{_{\rm exp}}^{^{|\Delta \epsilon|}}$. Therefore, with the weak measurement technique we measure the amplified difference between the CGH shift for the $s$ and $p$ polarizations, and not the individual CGH shifts for these two waves. We repeated this procedure for various angles of incidence around $\theta_c$ ($-5.80^\circ \leq \theta \leq -5.55^{\circ}$), always keeping the prism-camera distance constant, within experimental error. We investigated two different axial distances: $z = 20$~cm and $z = 25$ cm. Very near $\theta_c$, the two peaks have intensities that are nearly of the same magnitude, as seen in Fig.~2c. As a result, the technique became very inaccurate, preventing us from obtaining reliable data at $\theta \approx \theta_c$.

\section*{\color{darkred} \normalsize III. THEORY}
The weak measurement model of~\cite{2013OL} does not apply to the critical region. A modified theoretical model for optical weak measurements near $\theta_c$ is reported in~\cite{2015JO}. We include here a brief description of that model for completeness. After  propagating through the BK7 prism, the $s$ and $p$ transmitted Gaussian electric fields of the incident finite optical beam can be written as~\cite{2015JO}:
%
\begin{equation}
E^{(s,p)}(\theta,y,z) \propto \exp \left[-\, \left(y - y_{_{GH}}^{(s,p)}(\theta,z)\right)^{2} \Big{/} w^{2}(z)\right].
\label{eq:EoutSol}
\end{equation}
The beam shift $y_{_{GH}}^{(s,p)}(\theta,z)$ depends on the input polarization ($s$ or $p$), angle of incidence $\theta$, and axial distance $z$, and it accounts for both the angular and lateral GH shifts.  In deriving Eq.~\ref{eq:EoutSol}, it was assumed that, without loss of generality,  the  transmittance of the prism in the critical region is approximately one. It is also assumed that the phase difference between the $s$ and $p$ waves picked up upon reflection at the prism's hypotenuse was removed by the QWP/HWP set.

The optical weak measurement technique amplifies the GH shift by mixing the $s$ and $p$ waves. After passing through the polarizer (oriented at angle $\alpha$) and the analyzer (at angle $\beta$), the transmitted intensity,  measured at the CCD camera,  is given by
%
\begin{eqnarray}
I(\theta,y,z) &  \propto & \Bigg{|} \cos \alpha \, \cos \beta \, \exp \left[-\, \frac{\left(y - y_{_{GH}}^{(p)}(\theta,z)\right)^{2}}{w^{2}(z)}\right] + \nonumber \\  &  &
\sin \alpha \, \sin \beta \,  \exp \left[-\, \frac{\left(y - y_{_{GH}}^{(s)}(\theta,z)\right)^{2}}{w^{2}(z)}\right] \Bigg{|}^{2}.
\label{eq:EoutCombSimlp}
\end{eqnarray}
For $\alpha=\pi/4$ and $\beta=3\, \pi/4 + \Delta \epsilon$, after simple algebraic manipulations, we obtain
\begin{eqnarray}
 I(\theta,y,z) & \propto & \Bigg{|}\, (\Delta \epsilon + 1)\, \exp \left[- \left( \frac{y - \bar{y} - \Delta y_{_{GH}}(\theta,z) / 2 }{ w(z)}\right)^{^2} \right] + \nonumber \\ && (\Delta \epsilon - 1) \exp \left[- \left( \frac{y - \bar{y} + \Delta y_{_{GH}} (\theta,z)/ 2 }{ w(z)}\right)^{^2} \right]  \,\Bigg{|}^{^{2}},
\label{eq:eq2SPb}
\end{eqnarray}
where $\bar{y} = \left(y_{_{GH}}^{(s)} + y_{_{GH}}^{(p)}\right) / 2$ and $\Delta y_{_{GH}}= y_{_{GH}}^{(p)} - y_{_{GH}}^{(s)}$. We see from the above equation that the beam intensity at the CCD camera consists of two Gaussian peaks separated by $\Delta y_{_{GH}} (z)$. To determine $\Delta y_{_{GH}} (z)$, we must find the location of the intensity maxima of the two peaks. Expanding the exponential and using $\Delta y_{_{GH}} \ll w(z)$, we get
\begin{equation}
 I(\theta,y,z) \propto  \,\left[\, \Delta \epsilon + \frac{\Delta y_{_{GH}}(\theta,z)}{w^{2}(z)}\,(y - \bar{y}) \right]^{2}\, \exp \left[- \frac{2\,(y - \bar{y})^{2}}{w^{2}(z)}\right].
\label{eq:outintensity}
\end{equation}
The intensity maxima of Eq.~\ref{eq:outintensity} are found at
\begin{equation}
y^{\pm}_{_{\rm max}} (\theta,z) =  \bar{y} + \frac{- \Delta \epsilon \pm \sqrt{(\Delta \epsilon)^{2} + 2 \left[\Delta y_{_{GH}}^{2}(\theta,z) / w^{2}(z) \right]}}{2\,\Delta y_{_{GH}}(z)} \, w^{2}(z).
\end{equation}
To perform a weak measurement, we select two symmetric values of $\Delta\epsilon$, i.e.~$\Delta \epsilon=\pm |\Delta \epsilon|$. For  $\Delta \epsilon=+|\Delta \epsilon|$, the intensity peak at  $y^{+}_{_{\rm max}} (z,|\Delta \epsilon|)$  is greater than the one  at $y^{-}_{_{\rm max}} (z,|\Delta \epsilon|)$. For  $\Delta \epsilon=-|\Delta \epsilon|$, the main peak is located at  $y^{-}_{_{\rm max}} (z,-|\Delta \epsilon|)$. Experimentally, the measured quantity is the distance between these two main peaks:
\begin{equation}
\label{eq:DeltaYmax}
\Delta y_{_{\rm exp}}^{^{|\Delta \epsilon|}}(\theta,z)  =\frac{- \left|\Delta \epsilon\right| + \sqrt{\left|\Delta \epsilon\right|^{2} + 2 \left[\Delta y_{_{GH}}^{2}(\theta,z) / w^{2}(z) \right]}}{\Delta y_{_{GH}}(\theta,z)} \, w^{2}(z).
\end{equation}
\noindent
When the polarizer and analyzer are crossed ($|\Delta \epsilon|=0$),  then the transmitted intensity corresponds to two peaks separated by approximately $\Delta y_{_{\rm exp}}^{^{0}} \approx \sqrt{2}\,\, w(z)$. For $\theta$ sufficiently greater than $\theta_c$, the condition $|\Delta \epsilon| \gg \Delta y_{_{GH}} / w(z) \approx \lambda / w(z)$ is satisfied, and the weak measurement technique leads to the experimental quantity $\Delta y_{_{\rm exp}}^{^{|\Delta \epsilon|}}\approx  \Delta y_{_{GH}}/|\Delta \epsilon|$. The small beam shift $\Delta y_{_{GH}}$ is amplified by a factor of $1/|\Delta \epsilon| \gg 1$. This amplification of $1/|\Delta \epsilon|$ corresponds to the amplification factor of 100 in the paper of Merano and colleagues~\cite{2013OL}.

However, in the critical region, we have $\Delta y_{_{GH}} \propto \sqrt{\lambda \, w(z)} \gg \lambda$~\cite{2015JO}. Consequently,  the condition $|\Delta \epsilon| \gg \Delta y_{_{GH}} / w(z)$ is no longer valid.  To obtain $\Delta y_{_{GH}}$ from the experimentally measured $\Delta y_{_{\rm exp}}^{^{|\Delta \epsilon|}}$, we solve Eq.\,(\ref{eq:DeltaYmax}) without any further approximation, yielding
\begin{equation}
\Delta y_{_{GH}}(\theta,z) = \frac {\left|\Delta \epsilon\right|}{1 -
\displaystyle{\left( \Delta y_{_{\rm exp}}^{^{|\Delta \epsilon|}}(\theta,z) \Big{/} \left[ \sqrt{2} \,\, w(z) \right] \right)^{2}}} \, \Delta y_{_{\rm exp}}^{^{|\Delta \epsilon|}}(\theta,z).
\label{eq:Deltagh}
\end{equation}

\section*{\color{darkred} \normalsize IV. EXPERIMENTAL RESULTS}

The measurable values $\Delta y_{_{\rm exp}}^{^{|\Delta \epsilon|}}$ of the GH shift are necessarily a combination of the angular and lateral GH shifts. Therefore, $\Delta y_{_{GH}}(\theta,z)$ corresponds to the composite (differential) GH shift.

Equation~\ref{eq:Deltagh} leads to a new weak-measurement amplification factor
 \begin{equation}
{\mathcal A(\theta,z)} =  \frac {1 -
\displaystyle{\left( \Delta y_{_{\rm exp}}^{^{|\Delta \epsilon|}}(\theta,z) \Big{/} \left[ \sqrt{2} \,\, w(z) \right] \right)^{2}}}{\left|\Delta \epsilon\right|},
\label{eq:Afactor}
 \end{equation}
which for $\theta > \theta_c + \lambda/w_0$ reproduces the amplification factor $1/|\Delta \epsilon|$ of~\cite{2013OL}. The amplification factor $\mathcal{A}(\theta,z)$ is not a constant, but it depends on the measured $\Delta y_{_{\rm exp}}^{^{|\Delta \epsilon|}}(\theta,z)$. Therefore, the weak-measurement amplification factor depends on the incidence angle $\theta$ and the axial distance $z$.

Table~\ref{tab:tab1} shows our experimental data taken at $z = 20$~cm and Table~\ref{tab:tab2} the data for $z = 25$~cm. Both tables show the measured values $\Delta y_{_{\rm exp}}$, the amplification factors ${\mathcal A}$, and the CGH shift $\Delta y_{_{GH}}$. We see from the data in the tables that the amplification factor is not the constant factor of $1/|\Delta \epsilon| = 100$ expected for incidence angles far above $\theta_c$, but it varies with $\theta$ and $z$. Near $\theta_c$,  ${\mathcal A}$ is smaller than $100$, but it tends to that value the farther $\theta$ goes below $\theta_c$.

In Fig.~3, we show our experimental results for the CGH effect, after correction by the amplification factor of Eq.~\ref{eq:Afactor}, as a function of incidence angle $\theta$. In (a) and (b), the beam propagation distance was $z = 20$ and $25$~cm respectively, and $\Delta\epsilon = \pm 0.01$. The CGH shift $\Delta y_{_{GH}}$ at $z= 0$  (evaluated from Eq.~(39) of \cite{ana1}) is shown as a dashed line. From this solution, we see that the CGH peaks slightly above $\theta_c$ and goes to zero as $\theta$ decreases. A numerical solution for the CGH shift~\cite{2015JO}, which takes into account the nonlinear axial dependence, is shown as a solid line. The numerical solution predicts CGH shifts dramatically different (enhanced) from those at $z = 0$. Due to this enhancement, the CGH shift becomes appreciable for $\theta < -5.70^{\circ}$ for which it is negligible at the near field. This enhancement cannot be explained as a simple geometrical amplification that results from beam expansion since it would lead to an amplification factor constant in $\theta$. The numerical solutions are in very good agreement with most of the corrected experimental data, confirming the nonlinear axial dependence predicted in~\cite{1987JOSAA,2015JO} that results from the interplay of the angular and lateral GH shifts. The agreement also confirms the validity of the new amplification factor ${\mathcal A(\theta,z)}$. Due to physical constraints of the experimental apparatus, it was not possible for us to measure the CGH shift near the glass-air interface (in the near field). Small disagreements between theory and experiment can be attributed to the fact that while the numerical solution corresponds to position shifts of the beam's intensity peaks, the measured data are associated with centroid position shifts. Due to the resolution of our rotation mount, we were limited to taking data at angular steps of at least $0.05^{\circ}$. Although the prism legs were anti-reflection coated, measurement of the beam centroid position was still affected by multiple internal reflections inside the prism that inevitably superposed with the main beam causing small distortions in its intensity profile, affecting the measurement of the beam's centroid position. At some incidence angles, the distortions were significant, causing gaps in our experimental data at different $\theta$ for different $z$. In this sense, the data taken at $z = 20$~cm and $z = 25$~cm are complementary.

\section*{\color{darkred} \normalsize V. CONCLUSIONS}

In summary, we have experimentally observed the composite GH shift of a focused visible beam via a weak measurement scheme in the critical region. Our work extends the applicability of the weak measurement technique with regards to the study of the GH shift by experimentally verifying the new amplification factor proposed in~\cite{2015JO}, and it complements previous studies of the GH shift via weak measurements~\cite{2013OL,2014OLa}. Our experimental data confirmed the {\em nonlinear} axial dependence that results from the interplay of the angular and lateral GH shifts theoretically predicted in~\cite{1987JOSAA,2015JO}.

\vspace*{1cm}

\noindent \textbf{\footnotesize \color{darkred} ACKNOWLEDGEMENTS}\\
The authors thank the Funda\c{c}\~{a}o de Amparo \`a Pesquisa do Estado de S\~ao Paulo (FAPESP) and the Conselho Nacional de Desenvolvimento Cient\'ifiico e Tecnol\'ogico (CNPq) for the financial support.

\newpage

\begin{table}[htbp]
\centering
\caption{Measured data ($\Delta y_{_{\rm exp}}$), amplification factor ${\mathcal A}$, and CGH shift $\Delta y_{_{GH}}$ as a function of the incidence angle $\theta$ for $z=20$~cm and $\left|\Delta \epsilon\right| = \pm 0.01$. At $z = 20$~cm, $w(z) = 299 \, \mu$m.}
 \vspace*{0.8cm}
\begin{tabular}{cccc}
\hline
\rowcolor{bgA} $\theta\left(\pm 0.05^{\circ}\right)$ & $\Delta y_{_{\rm exp}}^{0.01}(\pm5) \,[\mu {\rm m}]$ & ${\mathcal A}$ & $\Delta y_{_{\rm GH}}[\mu{\rm m}]$  \\ \hline
 $-5.55^{\circ}$ & $360$ & 36 & $10\pm 1 $ \\
 $-5.70^{\circ}$ & $310$ & 56 & $5,5\pm 0.6 $ \\
 $-5.75^{\circ}$ & $280$ & 67 & $4,2\pm 0.5 $ \\
 $-5.80^{\circ}$ & $250$ & 77 & $3,3\pm 0.4 $ \\ \hline
 \end{tabular}
   \label{tab:tab1}
 \end{table}

 \begin{table}[htbp]
\centering
\caption{Measured data ($\Delta y_{_{\rm exp}}$), amplification factor ${\mathcal A}$, and CGH shift $\Delta y_{_{GH}}$ as a function of the incidence angle $\theta$ for $z=25$ cm and  $\left|\Delta \epsilon\right| = \pm 0.01$. At $z = 25$~cm, $w(z) = 352 \, \mu$m.}
 \vspace*{0.8cm}
 \begin{tabular}{cccc}
\hline
\rowcolor{bgA}   $\theta\left(\pm 0.05^{\circ}\right)$ & $\Delta y_{_{\rm exp}}^{0.01}(\pm5) \,[\mu {\rm m}]$ & ${\mathcal A}$ & $\Delta y_{_{\rm GH}}[\mu{\rm m}]$  \\ \hline
 $-5.55^{\circ}$ & $410$ & 41 & $10\pm 1$ \\
 $-5.65^{\circ}$ & $415$ & 39 & $11\pm 1 $ \\
 $-5.70^{\circ}$ & $355$ & 59 & $6.0\pm 0.7 $ \\
 $-5.80^{\circ}$ & $280$ & 80 & $3.5\pm 0.4 $ \\
\hline
 \end{tabular}
  \label{tab:tab2}
\end{table}

\newpage

\begin{figure}[htbp]
\vspace{-7.6cm}
\hspace*{-3.2cm}
\includegraphics[width=23cm,height=38cm]{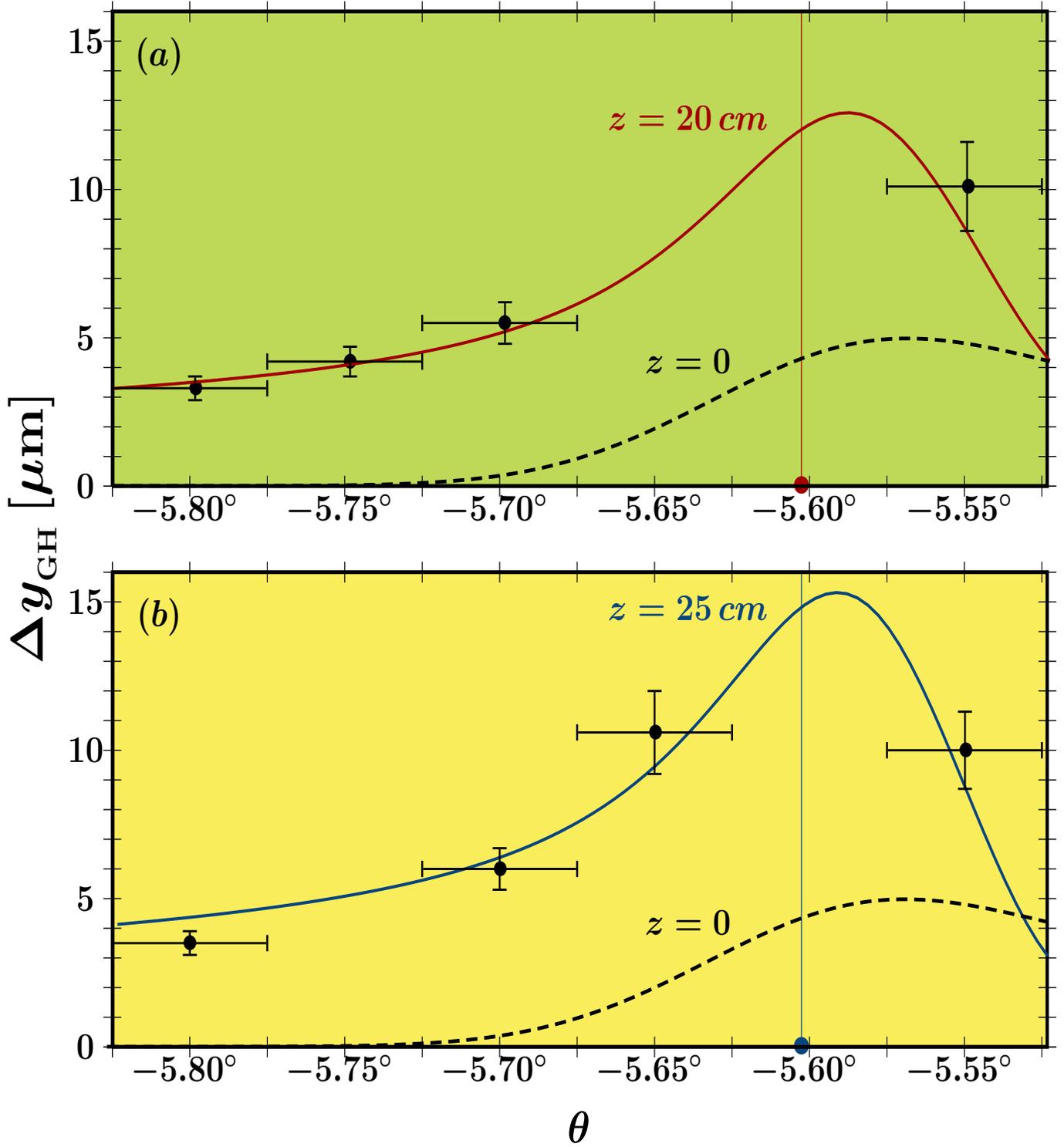}
\vspace*{-9.8cm}
 \caption{The composite (differential) GH shift $\Delta y_{_{GH}}$ in the critical region as a function of incidence angle $\theta$. The solid lines correspond to the calculated CGH shift at (a) $z=20$~cm and (b) $z = 25$~cm; the dashed line in both cases is the CGH evaluated at $z = 0$. The experimental results are represented by the solid circles with error bars. The vertical lines indicate the critical angle $\theta_c = -5.603^{\circ}$.}
\label{fig:Fig3}
\end{figure}

\end{document}